\documentclass[12pt,a4paper]{article}
\usepackage[top=25mm,left=25mm,right=25mm,bottom=20mm]{geometry}%
\usepackage[comma]{natbib}
\usepackage{footmisc}
\usepackage{amsmath,amssymb,array}
\usepackage{hyperref}
\usepackage{booktabs}
\usepackage{longtable}
\usepackage{multirow}
\usepackage[utf8]{inputenc}
\usepackage{csquotes}
\renewcommand{\mkbegdispquote}[2]{\itshape}
\usepackage{graphicx}
\usepackage{url}
\usepackage{lscape}
\usepackage[english]{babel}
\usepackage{commath}
\usepackage{amsfonts}
\usepackage{graphicx}
\usepackage{stackengine}
\usepackage{bm}
\hbadness=\maxdimen
\vbadness=\maxdimen
\usepackage{float}
\usepackage[nottoc]{tocbibind}

\begin{document}
	
\title{Six Years of \textit{Shiny} in Research - Collaborative Development of Web Tools in R}
\author{by Peter Kasprzak\thanks{University of Adelaide, School of Agriculture Food and Wine, SA 5064, Australia.} \thanks{peter.kasprzak@adelaide.edu.au}, Lachlan Mitchell\footnotemark[1] \thanks{lachlan.mitchell@icloud.com}, \and Olena Kravchuk\footnotemark[1] \thanks{olena.kravchuk@adelaide.edu.au}, Andy Timmins\footnotemark[1] \thanks{andy.timmins@adelaide.edu.au}}

\maketitle

\abstract{
\noindent
The use of \textit{Shiny} in research publications is investigated. From the appearance of this popular web application framework for R through to 2018, it has been utilised in many diverse research areas. While it can be shown that the complexity of \textit{Shiny} applications is limited by the background architecture, and real security concerns exist for novice app developers, the collaborative benefits are worth attention from the wider research community. \textit{Shiny} simplifies the use of complex methodologies for  users of different specialities, at the level of proficiency appropriate for the end user. This enables a diverse community of users to interact efficiently, utilising cutting-edge methodologies. The literature reviewed demonstrates that complex methodologies can be put into practice without the necessity for  investment in professional training. It would appear that \textit{Shiny} opens up concurrent benefits in communication between those who analyse data and those in other disciplines, thereby potentially enriching  research through this technology.
}

\section{Introduction}
Data is the backbone of research. With the rise of automated data gathering tools, data size and the complexity of analysis have driven a growing gap between research disciplines and the required data analysis. Another issue is the fact that different approaches to the same data can compromise validity, as seen in an analysis of effect sizes in observational studies, with varied methodological work-flows having the potential to reverse conclusions regarding the studied intervention \citep{donoho_50_2017}.  Collaborative learning employing common task frameworks can help interpret, quantify, and possibly cap methodological variations across disciplines \citep{donoho_50_2017}.  Software such as Matlab \citep{moler_matlab_2012}, Minitab \citep{arend_minitab_2010}, Genstat \citep{payne_genstat_2007} and SPSS \citep{landau_handbook_2004} have attempted to bridge this gap by creating more user friendly interfaces that either make coding more intuitive and easier to learn, or use drop down menus and radio button selection to bypass the command line. Current analytical software, such as those mentioned above, each have their own limitations that include non-publication ready quality graphics, unintuitive drop down menus, restrictive interfacing with other software, price point (including the cost of licensing the proprietary software) and the difficulties that inevitably occur when colleagues attempt to run code originating from other software on their preferred platform. Despite its own weaknesses, which include a very steep learning curve and unintuitive programming language, R has grown to become the most popular programming language for statistics and biological data analysis, spawning over 14,000 free-to-use packages over a wide range of subject material \citep{li_bioinstaller_2018}.

While code of any language can easily be shared between users, general use requires a level of familiarity with the specific program. Until recently, transforming a piece of R code into an interactive app capable of use by a broad audience required either knowledge of other coding languages, or consultation/collaboration with a computer scientist/app developer, requiring   justification  of the additional time and cost \citep{gunuganti_application_2018}.  Specialist apps that benefit only a small number of people  often do not meet cost/benefit benchmarks. This means that many useful advancements have stalled due to the requirement for experience with data analysis software, or an understanding of the underlying theory for day to day use, constituting a complexity barrier \citep{depalma_disk_2013}.  \textit{Shiny} can generalise R code for all levels of users, bringing the latest advancements in methodology measurably closer to everyone. This does create new issues relating to data security, as novice app developers will now require a knowledge of web internet protocols for secure data transfers to be assured.

The increased use of technology, sensors and other data capture devices has highlighted a new and interesting issue: researchers and practitioners without a background in data analysis now have the ability to gather large amounts of data \citep{lazerte_feedr_2017}, yet few options exist for those without data analysis training to analyse the ‘big’ data gathered from the field and experiments correctly. This has arguably led to issues impacting experimental reproducibility. A \textit{Nature} survey of 1,576 researchers from the disciplines of chemistry, physics and engineering, earth and environment, biology, medicine and other, found more than 50\% of those surveyed believed that low statistical power or poor analysis was a strong contributor to irreproducibility \citep{baker_1500_2016}.  In the same survey, more than 90\% of respondents believed that a better understanding of statistics was required to drive reproducibility of research. Learning analytical methodologies and programs is a non-trivial task, and subcontracted analysis, even in-house, generally comes with a wait for results. Purchasing proprietary software can be inflexible and often expensive, taking resources away from research, while open source software is dependent on a minimum level of computer literacy, along with the ability to test the software to ensure correct results \citep{lazerte_feedr_2017}.

Open source and free, \textit{Shiny} has grown in popularity with the first \textit{Shiny} Developer Conference held in January 2016 and growing use in peer reviewed academic papers. While the number of papers has steadily increased each year, \textit{Shiny} remains an incompletely explored topic, with the potential for \textit{Shiny} to make a significant positive contribution to the general field of science not yet properly examined. To the best of our knowledge this is the first \textit{Shiny} review.

The rest of this review is organised as follows. Section 2 details the literature search, the keywords and findings. Section 3 presents the technical aspects of \textit{Shiny}, including hosting costs and security, along with its restrictions. Section 4 discusses the use of \textit{Shiny} in research with relevant examples from the literature, and finally section 5 gives the conclusion along with the authors’ evaluations.

\section{Algorithms/methods for the literature search}
A thorough search for \textit{Shiny} results in the academic literature was undertaken to investigate the growth in research from 2012 - 2018, and the types of publications and subject areas represented. The focus of this paper is the use of \textit{Shiny} to bridge specialist academic and theoretical innovations, and their role in disseminating knowledge to governments, industry and the general community. Therefore, it is acknowledged that this is a non-exhaustive list of \textit{Shiny} case uses. We acknowledge that the literature search is not fully comprehensive, as newspaper articles, blogs and other non-academic areas were filtered.  This makes this review biased towards  an academic standpoint. The search was conducted with four major data bases. Web of Science and Scopus were used due to their reputation as multi-disciplinary databases,  with Google Scholar and the University of Adelaide (UofA) utilised  as their algorithms search the entire document and all fields for keywords. The keywords used in all searches were of the form, "\textit{Shiny} Web Application" OR "\textit{Shiny} Web App", with an exact search not suitable in this case, and "R" not included to avoid the inevitable non-related hits. The search was then filtered by year to span 2012 - 2018 and the document type was limited to Dissertations, Articles, Conference proceedings and Reviews (which were allowed), to investigate the use of \textit{Shiny} in the research literature only. Books were excluded from the search due to the small number of published materials. A separate search conducted for books showed that as of 2016, only two books were written on the use of \textit{Shiny}, with both being structured as instructional manuals.  \citet{beeley_web_2013} takes the beginner from their first application and walks them through the major concepts to more complicated applications, while \citet{moon_learn_2016} uses \textit{Shiny} to teach \textbf{ggplot2} \citep{wickham_ggplot2_2018} graphics. As of 2018, a Google search for "\textit{Shiny} Web Application Books" yielded seven results, including one second edition release.

\begin{figure}[htbp]
	\centering
	\includegraphics[width=\textwidth]{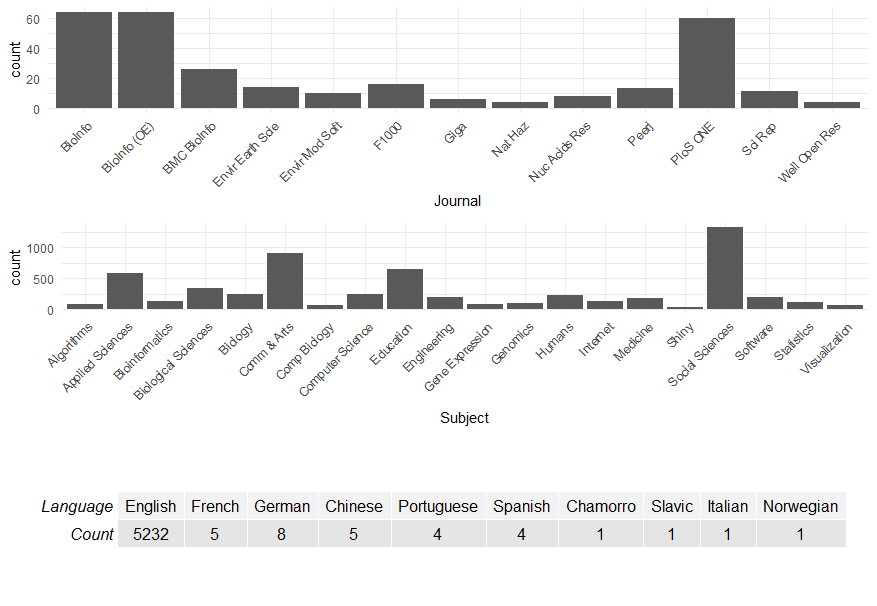}
	\caption{Summary of UofA search results partitioned into number of results by journal for records \(> 3\) with abbreviations given in \autoref{Tab:Abb}, results by subject tag with abbreviations given in \autoref{Tab:Abb2}, and published languages.}
	\label{Tab:fig_1}
\end{figure}

The UofA search engine is powered by ExLibris Primo and includes all resources owned or subscribed to by the library and selected free and open access resources. It includes 345 databases, and links to the major collections of articles and eBooks totaling over 50 million items, which can expand out to 100s of millions. The UofA search was conducted with the terms, "\textit{Shiny} web app OR \textit{Shiny} web application", and returned 5,251 results with 1,391 peer reviewed articles, 3,456 dissertations, 18 reviews and 119 conference proceedings. These are broken into results by journal title, subject tag, language published and are displayed in \autoref{Tab:fig_1}.

\begin{figure}[htbp]
	\centering
	\includegraphics[width=\textwidth]{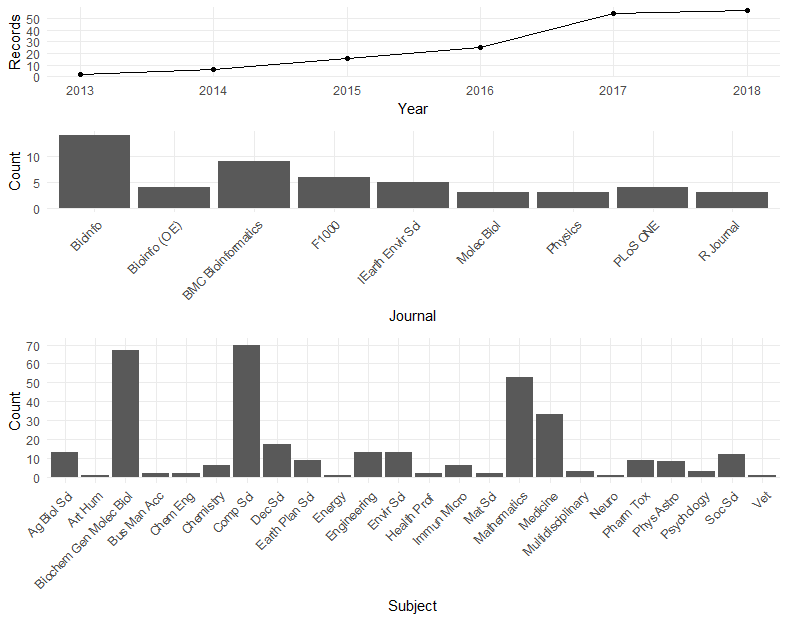}
	\caption{Summary of Scopus search results partitioned into number of published results by year, journal title for records \(> 2\) with abbreviations given in \autoref{Tab:Abb} and results by subject tag with abbreviations given in \autoref{Tab:Abb2}.}
	\label{Tab:fig_2}
\end{figure}

The search in Scopus used TITLE-ABS-KEY(\textit{Shiny} AND web AND app*) AND PUBYEAR \(>\) 2012 AND PUBYEAR \(<\) 2019 as its search terms, with the same document limitations.  The decision was made to check only the title, abstract and keywords, as too many irrelevant results were being returned when including other fields and gave a final result of 155 items. These were restricted once again to articles (114), conference papers (38), conference reviews (2), and reviews (1). These are broken into the number of records published by year, journal title and subject tag and are displayed in \autoref{Tab:fig_2}.  There were 154 records published in English, with one record published in Spanish.

\begin{figure}[htbp]
	\centering
	\includegraphics[width=\textwidth]{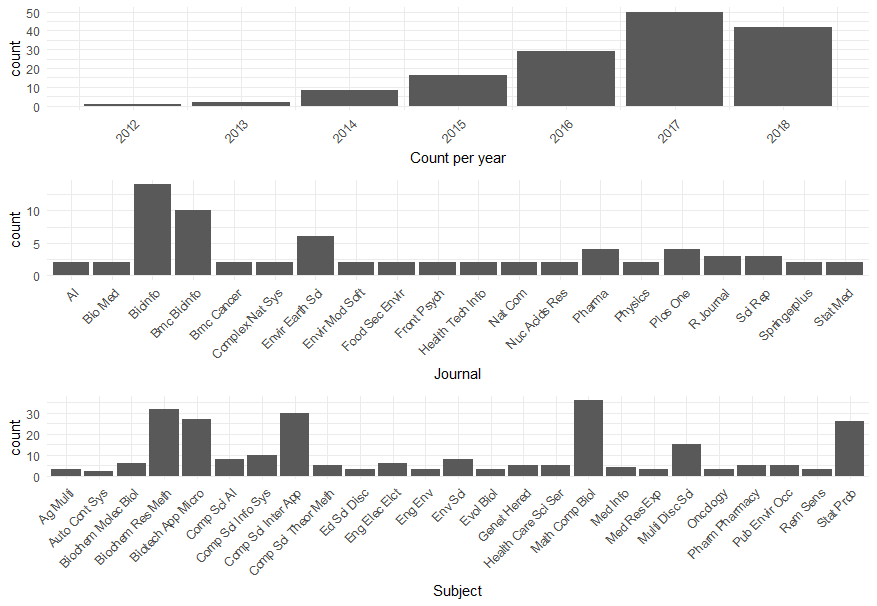}
	\caption{Summary of WOS search results partitioned into number of published results by year, journal title for records \(> 2\) with abbreviations given in \autoref{Tab:Abb} and results by subject tag \(> 2\) with abbreviations given in \autoref{Tab:Abb2}.}
	\label{Tab:fig_3}
\end{figure}

The Web of Science (WOS) search returned 144 results using the search criteria ALL=(\textit{Shiny} Web App*) and filtered to the same time frame.  Choices of document criteria included Articles and Proceedings papers which resulted in 110 Articles and 34 Proceedings papers, with dissertations not returned in this search.  These are once again broken into publications per year, journal title and subject tag displayed in \autoref{Tab:fig_3}.
Again the predominate language was English with 142 records, one Spanish, and one Portuguese record found.

The Google Scholar search terms used first were [Shiny web app | application] and returned 16,400 results. A range of additional terms were used to narrow down results including, "security OR complexity OR architecture OR hosting", with "Shiny" being a required keyword, and this search returned approximately 10,400 results. Unfortunately, 135 irrelevant results were found that did not contain the required term "Shiny".  This appeared to be an error in the algorithm. Given that the search returned over 100\% more results than the UofA search, it was decided to not use these results to create this paper, as the UofA search utilised the Google Scholar databases.
	
\textit{Shiny} is a relatively novel tool with the total number of papers found being relatively small. Assuming the UofA library search completely covers the other three databases (which it is advertised to do), there is an approximate total of 5,000 unique peer reviewed pieces of work utilising \textit{Shiny} since 2012; an average of over 700 papers per year. All searches showed that Bioinformatics journals published the largest number of \textit{Shiny} papers; however, the vast majority of papers were published by a diverse range of titles, in a diverse range of fields. This indicates that \textit{Shiny} is a flexible tool and not area specific. Only the Scopus search returned slightly different information with Computer Science, Biochemistry, Mathematics, and Other subject tags registering the largest number of relevant hits. On closer inspection, while Bioinformatics did not register as a subject heading, the journal that published the greatest number of papers was still Bioinformatics, followed by BMC Bioinformatics. This suggests that there is simply a difference in subject labelling. Far more papers were found by keyword searches in the body of the document, as evidenced by the total numbers of papers found by the library search from the UofA. This suggests that \textit{Shiny} has been utilised as a general tool, and not as a new discovery in the later years. The vast majority of all papers were written in English, with some European countries represented, and very few Chinese papers.
	
The results of the search algorithms are reasonably reproducible, with some fluctuations occurring depending on the sources of publications, and performance of the search engine. In our experience, the fluctuation is less than 10\%. Google Scholar significantly alters the number of found papers depending on sorting. If sorting by relevance is checked, then 127,000 results are found. Sorting by date reduces this number to what is stated above.

A subset of 600 papers was chosen for thorough reading to inform this report. These were the top 600 results returned by the UofA records search when sorted via relevance. The relevance ranking employed by ExLibris Primo is comprised of four main criteria:
\begin{enumerate}
	\item Degree of match: Fields such as title, author and subject field are given a higher ranking, along with the order of the query terms and completeness of phrases.
	\item Academic significance: Citations and journal impact factor.
	\item Type of search: Primo infers if the search is broad-topic or specific-topic, with broad topic searches amplifying overview material such as reference articles.
	\item Publication date: Newer material is given preference.
\end{enumerate}
The papers that discussed \textit{Shiny} generally had "\textit{Shiny}" in the title and/or the subject fields, increasing their relevance score. Earlier papers were more likely to discuss \textit{Shiny}, with newer papers more likely to mention \textit{Shiny} in the text only. The relevance search yielded a high number of the older papers as high relevance, along with a very broad range of use cases. The limit of 600 papers was an empirical cut off point, as this was the stage at which papers had ceased discussing \textit{Shiny}, and were only stating its use. It was decided that enough use cases had been examined to make comments regarding \textit{Shiny}’s relatively widespread use in the body of academic work. 445 original applications were introduced in these papers, which utilised 373 unique R packages. 229 unique peer reviewed journals were represented, with 55 published in Bioinformatics, 31 published in PLoS ONE, and 21 published in BMC Bioinformatics. The final subset of papers that most thoroughly discussed the implementation of \textit{Shiny} were chosen to create this report and are given as references.

\section{Technical aspects}
\subsection{Architectural overview}
A Web application framework for R, \textit{Shiny} was conceptualised by RStudio's CTO Joe Cheng and announced at the Joint Statistical Meeting conference in July of 2012 as a tool designed to help R programmers create interactive web applications, reports and analysis without the need to know HTML, CSS, or JavaScript \citep{chang_shiny_2018}.

The power of \textit{Shiny} comes from its ability to enable an R user to quickly and simply code a reactive framework. A reactive framework allows objects to be updated when a source is changed, along with all connected objects. For example, in an imperative programming paradigm such as R language, setting the line:
\[c = a + b\] 
means that \(c\) is assigned the sum of previously defined terms \(a,b\) and will not change when the values of \(a,b\) are changed without the variable \(c\) being re-evaluated.  Reactive programming allows the value of \(c\) to be updated almost instantaneously, including all other variables and outputs dependent on \(c\), whenever \(a\) or \(b\) is changed.  R completes this task with information travelling from input to output in a pull fashion.  A pull fashion is when \(c\) learns of the new value of \(a\) or \(b\) when \(c\) is called.  \textit{Shiny} creates a system of alerts that flag changed expressions and the server re-evaluates all flags in an event known as a \textit{flush} \citep{grolemund_shiny_2015}.  Using two object classes called reactive values, such as a = reactive(), and observers, such as b = plot(), \textit{Shiny} creates a \textit{reactive context} between the two objects known as a \textit{call-back} which is a command to re-evaluate the observer.  Multiple observers can be linked to the same \textit{reactive value} and the server will queue up all \textit{call-backs} and run each \textit{call-back} in the event of a \textit{flush} \citep{grolemund_shiny_2015}.

This reactive framework allows user inputs to be evaluated via a user interface (UI) with a series of easily-coded widgets, such as text boxes, radio buttons and drop down menus, from pre-programmed R code. \textit{Shiny} then seamlessly updates outputs of tables, plots and summaries. A non-R user can change the values of \(a\) and \(b\) via the user interface and explore the pre-coded results dependent on \(c\).

A \textit{Shiny} application has two main parts. A user interface object and a server function. The user interface contains code for the layout and appearance of the app, with default choices restricted in appearance. Layouts can be customised and changes to the appearance can be made if the programmer has some knowledge of HTML or CSS. For standard applications, simple commands suffice and a knowledge of HTML or CSS languages is not required for tweaks. The server function houses all the code that drives the functionality of the application and can utilise all the built-in programs available to R and RStudio users.

\subsection{Hosting}
For a small number of applications and limited run hours, the cost of hosting a \textit{Shiny} application is free, but it can become expensive quickly.  Hosting on shinyapps.io requires no system administration knowledge, and comes with layers of security and is supported by \textit{Shiny}'s IT team.  According to the RStudio pricing website \citep{core_team_rstudio_2012}, the platform is free for 5 applications and 25 active hours, then increases to \$39 AUD a month for unlimited applications and 500 active hours.  The top tier of \$299 AUD a month allows for unlimited applications and 10,000 active hours.  \textit{Shiny} also has the option of \textit{Shiny} server, \textit{Shiny} Server Pro or RStudio Connect.  These require a level of system administration knowledge, and also require the apps to be hosted on a physical or virtual machine.  RStudio Server Pro costs \$9,995 AUD per year \citep{core_team_rstudio_2012}.  RStudio connect allows installation of software on a server behind an existing firewall and costs between \$14,995 AUD per year (\$62 AUD per user/month) and \$75,995 AUD per year (\$6.25 AUD per user/month) for a larger, specified number of named users \citep{core_team_rstudio_2012}.  \textit{Shiny} server prices were not available at the time of going to print. For those with an in-depth knowledge of internet security, it is possible, and more economical, to host the application independently by their own means.

\subsection{Security}
As \textit{Shiny} is primarily a web technology, a very strong focus on application security must be adhered to, with novices in computer science more likely to make critical mistakes \citep{charpentier_web_2013}.  A well-known concept in cryptography and web security is \textit{unknown unknowns} \citep{charpentier_web_2013}.  Put simply, this refers to the fact that a developer cannot build defences for attack vectors with which they are unfamiliar. For this reason, it is generally wise to leave the specifics of data security to experts in the field, with the end-developer instead relying on the vetted work that has been done for them.

For users of \textit{Shiny} who elect to use shinyapps.io by RStudio, this is essentially what happens. Once uploaded, the application is secured behind best practices \citep{core_team_rstudio_2012}. Unfortunately, this service is prohibitively expensive when compared with hosting the server on a cloud platform like Amazon Web Services (AWS) \citep{amazon_amazon_2019} or Microsoft Azure \citep{microsoft_pricing_2019}.  This requires application security to be taken into the app creator's hands. Certificates need to be created and kept up-to-date, and servers need to be configured for HTTPS amongst other security protocols \citep{charpentier_web_2013}.  Due to the local nature of R, this is likely to be a new issue, requiring a new set of skills, for many data analysts operating on the platform.

\subsection{Architectural issues}
Curiously, there are only a small number of papers that explicitly mention concerns and limitations with respect to the use of \textit{Shiny} to develop research-focused apps.  A paper by \citet{dwivedi_shinygispa_2018} was the first to include a limitations section, emphasising the requirement for a fast internet connection when dealing with large data sets.  This could be mitigated with the use of cloud-based resources to store the data and host the app, with potentially faster network and processing speeds available with respect to local connections. \citet{guo_developing_2018} found R package updates a legitimate concern, as updates can occur without warning and crash an application. A less serious issue is the lack of flexibility of the dashboard, which derives from the simplification of its creation, with \textit{Shiny}’s dashboard not being as flexible as one created in Java \citep{ge_idep_2018}.  There are, however, challenges with the use of \textit{Shiny}, with one of them being the background architecture.

While \textit{Shiny} has many benefits, the architecture of \textit{Shiny} is a limiting factor when building complex applications. Previously this concern has been dismissed, with Joe Cheng stating more recently:
\begin{displayquote}
	In the past, we’ve responded rather glibly to these requests. Just use functions! \citep{cheung_web_2017}
\end{displayquote}
As of 2017, \textit{Shiny} has made moves to address this issue with the creation of modulisation \citep{cheng_shiny_2017}; however, more involved use cases would be handled better by other computing languages, for the reasons detailed below. An analogous way of conceptualising this would be in the difference between applets and applications. Applets are generally small, discrete, and of low complexity, and are developed to perform a small number of functions for a highly specific purpose: most \textit{Shiny} products would fit this description quite well.  Meanwhile, applications are generally more complex \citep{fayram_functional_2011}. They are built for a number of different use cases, and tend to have relatively large codebases. Well-established and popular web application frameworks such as Angular and React exist to fit these situations, containing much more general functionality than \textit{Shiny}, with much less specific functionality (such as functions related to data visualization) \citep{mitchell_shiny_2018}. None of this is to say that complex applications cannot be created with \textit{Shiny}, but it may not be the most mature solution for the task.

While \textit{Shiny} will undoubtedly continue to evolve in much the same way as R, and many issues today will be overcome tomorrow, a number of well-established software development paradigms must be diverged from:
\begin{itemize}
	\item \textit{Shiny} actively encourages the use of single-file applications, generally referring to this singular file as \texttt{app.R} \citep{core_team_shiny_2017}.  Defining everything in a singular file works well for prototypes, but quickly falls apart as an application grows and increases in complexity.  In general, code is compartmentalised into files which contain the logic for a single component.  By allowing a single file to grow monolithic in size, code readability and re-usability is challenged, consequently making it harder to add additional components in the future \citep{fayram_functional_2011}.
	\item \textit{Shiny} insists on a reactive data-driven model over the more traditional and common event-driven model.  While not necessarily a flaw in and of itself, many novice developers consider reactivity in programming to be a non-trivial concept \citep{fayram_functional_2011}. Considering that \textit{Shiny}, by nature, is aimed towards data analysts rather than computer scientists, it can increase the initial difficulty hurdle that beginners have to overcome. For example, a bug has existed in RStudio since at least February 2018, which prevents automatic reloading from working with sourced files. When using multiple files like this, the server needs to be stopped manually and restarted between every change, making for a tedious development cycle. Concerns on the subject have not been addressed by either the RStudio or \textit{Shiny} core developers \citep{hansen_rstudio_2018}.
\end{itemize}
Both of the above points begin to cause major issues when put together.  Encouragement of singular source files results in code quickly becoming unruly, threatening flexibility.  This heightened complexity of source code will invariably be replicated within the reactive dependency graph, \textit{Shiny}'s internal mapping of reactive nodes and their relationships.  In the event that something is not working as expected, RStudio provides little to no internal tools for debugging this graph.  A new addition to CRAN in the form of \textbf{reactlog} \citep{schloerke_reactlog_2019} is a first attempt to address this issue, which usually forces the developer to painstakingly debug the graph by hand. As the application becomes increasingly complex, this process gets closer and closer to impossible. Many of these cases remain undocumented as \textit{Shiny} is a burgeoning technology and, to the best of our knowledge, this paper is the most in-depth look at the challenges in the peer-reviewed literature.

A package, \textbf{ShinyTester} \citep{kohli_shinytesterfunctions_2017}, was added to CRAN \url{https://cran.r-project.org/} early in 2017.  While it provides a promising first approach to debugging tools for \textit{Shiny} (such as the inclusion of a dependency graph visualiser), it unfortunately seems to have been abandoned.  Tools like this would likely alleviate the above-outlined concerns.

\subsection{Data size}
\textit{Shiny} is designed foremost as a server technology, with applications intended to be used remotely with a stable internet connection \citep{core_team_shiny_2017}.  \textit{Shiny} applications must be built with upload and download requirements at the fore. While \textit{Shiny} applications can be run locally, doing so requires a base level of knowledge of R that may make it a sub-optimal approach, and limits accessibility. This is comparable with how mobile applications are generally shipped as pre-compiled binaries, rather than as raw sources that the user would need to compile and install manually. One of the largest issues with this inherent reliance on connectivity is the need for data to be uploaded and downloaded. Since \textit{Shiny} has no in-built data streaming functionality, it is not possible to work with parts of the data while waiting for the rest to upload \citep{core_team_shiny_2017}. An entire transfer must be completed before the dataset is made available to the application. This forces the application to require pre-partitioned uploads, which may not be possible for all types of datasets.

It is quite common to see a dataset approaching gigabytes in size, especially prevalent in areas such as genomic sequencing. It is generally technically unrealistic for datasets of this size to be worked with remotely, and would include extra data costs. If a large amount of bandwidth were made available to a single user, this could open up the user’s service to potential denial-of-service attacks by malicious entities  \citep{cloudflare_what_2019}.  Furthermore, it may be legally unrealistic in terms of data ownership. Users are often uncomfortable about providing sensitive data to unknown receivers, as there is no way for a \textit{Shiny} app to prove that it is not storing uploaded information permanently for the developer’s own academic or financial gain \citep{kacha_overview_2018}.

\section{Literature analysis}
\subsection{Complexity barrier}
The pattern of peer reviewed work, as shown in \autoref{Tab:fig_1}, \autoref{Tab:fig_2}, and \autoref{Tab:fig_3}, shows that Bioinformatics is a popular and growing area for \textit{Shiny} apps.  Areas that traditionally have a lower focus on data analysis skills for researchers, such as Biological Sciences, Education and Index Medicus, appear to have higher usage levels.  In the current literature, \textit{Shiny} is primarily used as a delivery/visualisation tool and so it is not the focus; with many papers referencing the use of \textit{Shiny} but not discussing its merits. This trend becomes obvious in more recent papers, with much of the best discussion occurring in earlier papers.

To investigate the uptake of \textit{Shiny}, we must first understand some of the factors that determine the uptake of innovation. These are stated by \citet{rogers_evolution_2001} as: (a) relative advantage, (b) compatibility, (c) complexity, (d) trialability and (e) observability.  \citet{rogers_evolution_2001} defined complexity as:
\begin{displayquote}
	...the degree to which an innovation is perceived as difficult to understand and use.
\end{displayquote}
The first peer-reviewed \textit{Shiny} publications appeared in 2013, with the first two dissertations contributing the most to this discussion, as they give a glimpses of \textit{Shiny}’s vast potential. The first dissertation using \textit{Shiny} was published by  \citep{depalma_determination_2017}.  To drive innovation and uptake, tools must be accessible and usable by all interested parties \citep{jahanshiri_developing_2014,klein_webxtreme_2017}.  \citet{moraga_spatialepiapp_2017} noted that the area of public health, while there had been progress in methodology and analysis:
\begin{displayquote}
	...these methods are still inaccessible for many researchers lacking the adequate programming skills to effectively use the required software.
\end{displayquote}
The first peer-reviewed \textit{Shiny} publications appeared in 2013, with the first two dissertations contributing the most to this discussion, as they give a glimpses of \textit{Shiny}’s vast potential. The first dissertation using \textit{Shiny} was published by  \citet{depalma_disk_2013}.  The \textit{Shiny} app allowed non-computer literate clinicians the ability to harness powerful statistical methodologies in a robust framework, and to conduct antimicrobial susceptibility tests to determine an unknown pathogen’s susceptibility to various antibioticss.  \citet{depalma_disk_2013} noted that previously new methods have not been adopted due to:
\begin{displayquote}
	...various computational difficulties and an absence of easy to use software for clinicians.
\end{displayquote}
Complex methodologies were able to be immediately used by end users without an assumption of computational skills, to inform important medical checks. This direct transfer of method is a concrete example of how \textit{Shiny} is able to make complex research available to all interested parties, regardless of knowledge level. Specialised applications such as this would be difficult and costly to create without \textit{Shiny} and, without general use software, any advanced methodology could stall in uptake due to complexity barriers. This was later followed up with dBETS (diffusion Breakpoint Estimation Testing Software) by  \citet{depalma_determination_2017}, who once again acknowledged:
\begin{displayquote}
	...the computational complexities associated with these new approaches has been a significant barrier for clinicians.
\end{displayquote}
\textit{Shiny} is a potential solution to the barrier of complexity for the uptake of new methodologies.  

\subsection{Cross-collaboration and dialogue}
Cross-collaboration between researchers and the easy dissemination of results is key to external validity \citep{munafo_manifesto_2017}.  \textit{Shiny} promotes collaboration by allowing people with varying skill levels access to more complex methodologies. This has a flow-on benefit to promote the collaboration of practitioners with researchers,  or field researchers with theorists, in order to create specialised, fit for purpose applications. This is illustrated in a paper by \citet{wages_web_2018}, which designs and conducts Phase 1 dose-finding trials using the continual reassessment method, and was noted to:
\begin{displayquote}
	...facilitate more efficient collaborations within study teams.
\end{displayquote}
\citet{klein_webxtreme_2017} underscores the requirement that
\begin{displayquote}
	...facilitating the deployment of web applications for data analysis is important to promote collaboration within the scientific community and between scientists and stakeholders.
\end{displayquote}
Further examples of \textit{Shiny} being used to open discussion by using apps to bring relevant parties with differing skills sets into collaboration include \citet{diaz-gay_mutational_2018}, who stated:
\begin{displayquote}
	...analysis of somatic mutational signatures remains currently inaccessible for a substantial proportion of the scientific community.
\end{displayquote}
as well as \citet{whateley_web-based_2015}, who noted the knowledge gap between relevant parties:
\begin{displayquote}
	...demonstrates the use of the \textit{Shiny} web framework to bridge that gap, allowing for collaborative development of web tools that can be coded in the widely-used and free R statistical computing language.
\end{displayquote}
The ability to bridge the gap between researchers and the tools required for their data analysis was mentioned by \citet{chen_interactive_2018} in the context of environmental DNA.  eDNA is becoming an essential tool in ecology and conservation biology and is utilised by a range of people with varying skill levels, with \citet{kandlikar_ranacapa_2018} stating:

\begin{displayquote}
	Results from eDNA analyses can engage and educate natural resource managers, students, community scientists and naturalists, but without significant training in bioinformatics, it can be difficult for this diverse audience to interact with eDNA results.
\end{displayquote}
\textit{Shiny} allows discipline specialists outside of computer science to code their own apps, bridging the skill gap for other researchers \citep{niu_mass_2017}.  This was demonstrated by an app called Armadillo Mapper \citep{feng_armadillo_2017}, which was designed specifically to decrease the time between synthesising distributional knowledge on a computer and carrying out conservation efforts in the field. This encourages those without the resources to conduct their own analysis to collaborate closely with analysts to create specialist applications. Rather than sending final data to an analyst for analysis, discussion and collaboration is encouraged at the beginning of an experiment. This sets up low quality data due to issues such as pseudo-replication, low power and confounding variables being avoided at the design stage rather than the analysis stage.

\subsection{Flexibility to link other software}
\textit{Shiny} has the flexibility to bridge the gap between specialised data gathering tools and available software.  A \textit{Shiny} app accompanying the R package \textbf{rHyperSpec} \citep{laney_toward_2013} was created to take complex data generated by hyperspectral cameras and link the data to available software packages in response to the problem of there being:
\begin{displayquote}
	...few free, open-source software packages that enable researchers to easily process and analyse such data in a manner that maximizes inter-comparison between studies.
\end{displayquote}
This showcases the flexibility of \textit{Shiny} applications being able to upload information in various formats, make appropriate changes, and output the data in a form usable by another, completely independent piece of equipment/software. Previously, there were precious few options to link independent software/equipment, especially without breaching warranty restrictions. \textit{Shiny} shows tremendous flexibility in working with existing infrastructure to help decrease costs, especially when technologies are in their infancy.

\textit{Shiny} gains flexibility and customisability directly from R. One of \textit{Shiny}’s most useful abilities is to wrap both existing and new R packages for general consumption.  \citet{beck_investigating_2014} created an app called \textbf{Seed}, which bundled several R packages together and used \textit{Shiny} to host them on the web, allowing:
\begin{displayquote}
	...user's access to powerful R based functions and libraries through a simple user interface. 
\end{displayquote}
In the Precision Agriculture (PA) space, farmers have access to a multitude of proprietary sensors, few of which can be linked directly to analysis tools \citep{jayaraman_internet_2016}.  \textit{Shiny}'s highly customisable framework facilitates the linking of several pieces of independent software, and can avoid manual data wrangling and transfer. As an example, \citet{jahanshiri_developing_2014} took data from existing PA sensors and utilised R functions for its analysis and visualisation of results.  \textit{Shiny} has proved more than useful in the results visualising area, with packages such as \textbf{\textit{Shiny}Stan} \citep{gabry_shinystan_2018} created in order to visualise modelling parameters and results from MCMC simulations.

\subsection{Generalising complex methodologies}
R packages can be thought of as a level of abstraction down from mathematical theory, as the packages can be used without the need to have full understanding of the methodology. \textit{Shiny} can be thought of as another level of abstraction down again, as R packages can be utilised without needing an understanding of R itself. This ability to generalise analysis methodologies makes \textit{Shiny} available to any interested party, driving flexibility, dialogue and cross-collaboration.

There is an overarching requirement when making tools available to a broader audience to ensure correct methodology.  The first example of using \textit{Shiny} to guide and educate the user came from \citet{assaad_rapid_2014} who created two \textit{Shiny} apps intended to allow Microsoft Word users access to One Way Anova analysis and post hoc tests.  The app gave instructions to guide users through the process, which greatly simplified the common statistical test, whilst promoting proper statistical methodology.  A real-world example of protecting the end user comes from \citet{hsu_attentional_2018}, who created an app for proper randomisation when allocating participants to a three-armed, double-blinded, randomized controlled trial (RCT) for depression.  One critical characteristic of the app was to ensure mistakes were not made when properly balancing strata.  A fail-safe against experimental error was employed by not allowing participants to have their experimental ID overwritten.  This means that any accidental changes after treatment has begun would not impact on the treatment received.

Generalised applications must be flexible to differing individual parameters. \textit{Shiny} makes it a simple task to allow parameters of a methodology to be changed, depending on individual circumstances. \textit{Shiny}-wrapped simulations were used to explore humanitarian responses and financial institution resiliency for earthquake risks in Indonesia, with  \citet{hartell_earthquake_2014} allowing the simulation to be tweaked by individuals so that adjustments to calibration parameters could be made, based on specific interests or circumstances.  Other apps that allowed the user to specify parameters were created by \citet{zhou_robustly_2014} for detecting differential expressions in RNA sequencing and \citet{yin_bayesian_2014}, who utilised Bayesian statistical modelling to investigate the networks of epidemic transmission.

\textit{Shiny} makes complex methodologies accessible to those who would previously not be part of the conversation; most likely due to a lack of theoretical study, or lack of familiarity with coding or analysis programs. \textit{Shiny} was explicitly noted to help increase engagement by \citet{lazerte_feedr_2017}, who created FeedR in order to record and visualise RFID data from ecological studies.  The huge amount of data from RFID quickly becomes overwhelming and requires specialist  methods to cope.  The FeedR \textit{Shiny} app was created to wrap the paired R package in order that:
\begin{displayquote}
	...this framework will become a meeting point for science, education and community awareness...we aim to inspire citizen engagement while simultaneously enabling robust scientific analysis \citep{lazerte_feedr_2017}.
\end{displayquote}

\subsection{Responsible and open research}
Reproducibility of research is a critical cornerstone of responsible research practices.  Studies, such as \citet{munafo_manifesto_2017}, have indicated that reproducibility is not at high enough levels, while results of a survey conducted by \citet{baker_1500_2016} and published in \text{Nature} found:
\begin{displayquote}
	...more than 70\% of researchers have tried and failed to reproduce another scientist's experiments and more than half have failed to reproduce their own experiments.
\end{displayquote}
Eight practices are argued for by \citet{munafo_manifesto_2017}, which include promoting transparency and open science to increase reproducibility. Open source software such as \textit{Shiny} can aid these objectives by creating a vessel to preserve code and allow a greater number of interested parties to critically evaluate methodologies and results.

One benefit of \textit{Shiny}-wrapped code is that methodology comparisons become much easier to conduct. Methodologies wrapped in \textit{Shiny} applications can be compared on a known data set under various conditions by the end user. This is a powerful tool in the advancement of reproducible research. \textit{Shiny} was explicitly used in a dissertation by  \citet{parvandeh_epistasis_2018} as a vessel to show this particular strategy and to create reproducibility of results, enhancing responsible and reproducible research goals.  

A \textit{Shiny} app, or at least the code behind it, is enduring.  A paper from \citet{sieriebriennikov_ninja_2014} included a \textit{Shiny} application named Nematode Indicator Joint Analysis (NINJA) 2.0, to automate manual calculations previously carried out using spreadsheet software, which is time consuming and prone to errors.  The aim for NINJA to remain freely accessible was validated when it was later used by \citet{burkhardt_perennial_2019} to aid nematode calculations in semi-arid wheat systems, five years after its release.  This suggests that maintaining a \textit{Shiny} application is not overly difficult.  The benefit of \textit{Shiny}'s easy maintenance and updating was mentioned for the first time in a dissertation by \citet{niu_mass_2017}, which highlighted the fact that only the source code requires changing, without having to download patches or modify individual applications. 

\textit{Shiny} also appeared in conjunction with machine learning to explore early phase drug discovery processes \citep{korkmaz_mlvis_2015}, with \citet{wojciechowski_interactive_2015} noticing the power of \textit{Shiny} to disseminate the results of research, stating:
\begin{displayquote}
	Interactive applications, developed using \textit{Shiny} for the R programming language, have the potential to revolutionize the sharing and communication of pharmacometric model simulations.
\end{displayquote}
Free and open source software is ideally suited to disseminating the products of research \citep{lazerte_feedr_2017}.  This also drives collaboration and was noted to encourage local and direct monitoring of environmental data in Kenya \citep{mose_application_2017}.  \citet{lazerte_feedr_2017} also found that \textit{Shiny}’s open source nature has another important benefit, which is that it:
\begin{displayquote}
	...reduces financial barriers to its use and the open-source aspect permits and encourages collaboration which can result in better, more powerful software.
\end{displayquote}
Cross-collaboration and use of open source \textit{Shiny} will hopefully also help to drive data sharing.  \citet{yi_zika_2017} noted the utility and importance of data sharing promoted by \textit{Shiny} applications.  This outcome is one of the key recommendations by \citet{munafo_manifesto_2017} in order to drive transparency and openness and is currently a policy for \textit{Science} and \textit{Springer Nature} journals.

\subsection{An educational tool}
The strengths shown by \textit{Shiny} seem to fit very well in the educational sector. Thus, \textit{Shiny} has frequently been used as a teaching aid to convey complex ideas to students \citep{williams_using_2018}.  Educational tools, such as those by \citet{arnholt_using_2018}, help teach the concept of power in hypothesis tests, with \citet{williams_using_2018} creating a similar application for confidence intervals.  An app by \citet{courtney_dealing_2018} normalises large datasets and enables students to explore the results of differing transformations.  There are other benefits to using \textit{Shiny} in the education sector.  \citet{kandlikar_ranacapa_2018} created the \textit{Shiny} app ranacapa and found that:
\begin{displayquote}
	A key benefit of using ranacapa was that, despite having no prior bioinformatics experience, students could begin exploring the biodiversity in their samples in a matter of minutes by using the online instance of the \textit{Shiny} app.
\end{displayquote}
This had the flow-on effect of enabling teachers to focus more on theory, instead of the inevitable problems when teaching new, more complex software, and provided a useful aid for self-learning  \citep{kandlikar_ranacapa_2018}.

\section{Conclusion}
This review examined \textit{Shiny} in peer reviewed publications from 2012 to 2018 and mapped the growth in its use through various research fields. A subset of 600 papers were used to inform the bulk of the paper, with the authors’ personal experiences of \textit{Shiny} included. While \textit{Shiny} is not a \textit{silver bullet} solution to issues in the research field, it confers the ability for specialised applications to be created cheaply and easily, such that any level of end user maybe included, no matter the complexity level of the methodology.  This primary benefit creates a direct pathway for new findings to be rapidly incorporated into established workflows. The flexibility of \textit{Shiny} means that apps can be tailored to exact specifications in all regards, with changes and maintenance of the app made relatively easy as an ongoing product of consultation, thereby further promoting collaboration. If an app is considered worthwhile adopting for an existing workflow, widespread adoption across an entire workplace is as simple as sharing the web address. This will have the inevitable effect of allowing fewer people to achieve more. This does necessarily mean existing jobs risk becoming obsolete: as new, potentially variant jobs will be created as a result of the new work undertaken. As we progress further into this technological world, this argument will require mature debate and nuancing to be resolved.

In the current literature, \textit{Shiny} has been used primarily as a visualisation and dissemination tool, with few papers exploring the concurrent benefits and challenges mentioned in this review. One benefit identified in the literature is the opportunity to increase high value dialogue between people with different skill sets. For example, field researchers, primary producers or marketers are able to work alongside theoretical researchers/consultants to create highly customised applications for up-coming experiments or daily work. Code published as a \textit{Shiny} application has the useful attribute of making methodology comparisons easy,  which promotes reproducible research and best practice standards.  

With the ability to accelerate access to data analysis techniques comes the paramount issue of data security for those unfamiliar with web protocols. It is essential for those who host web-based applications to become knowledgeable in this area. Web security protocols are likely to be a new skill set for many R programmers: a crucial task, potentially preventing widespread uptake of  \textit{Shiny}.

While the use of \textit{Shiny} apps requires minimal experience with computers, the creation of a \textit{Shiny} application is a different story. The lack of debugging tools, the encouragement of single file applications, and the current implementation of the reactive data-driven model will limit the complexity of future applications. Other open source and proprietary options are currently available, however, \textit{Shiny}’s flexibility, customisability, and low cost is highly desirable. Open source software comes with a minimum knowledge requirement barrier to entry, and proprietary software can be expensive and inflexible to changing situations and circumstances. Maintenance is required with \textit{Shiny}, although it is limited to updating code when R packages or dependencies change. This updating can be achieved via the source code for all users.

\textit{Shiny} is one of the better tools available if one is an existing R programmer, given its inherited scope from R. It helps promotes open and reproducible research and offers a real pathway for making complicated methodologies usable to those outside universities and supported research facilities. The ability to provide an avenue to increase high value collaboration and dialogue between interested parties with differing skills sets makes \textit{Shiny} a tool well worth exploring.

\section{Acknowledgements}
We would like to gratefully acknowledge the first author’s scholarship for the M.Phil program of the first author from the Grains Research and Development Corporation (GRDC) Australia.
\bibliography{kasprzak}
\newpage
\section{Appendix}
This appendix contains the table of abbreviations for journal titles and subject headings.\\

\setlength{\extrarowheight}{0.5cm}
\begin{table}[H]
	\caption{Table of journal abbreviations}
	\label{Tab:Abb} 
	\renewcommand{\arraystretch}{0.5}
	\centering
	\begin{tabular}{ p{12cm} | r }
		\toprule
		Journal & Abbreviation \\
		\toprule
		2nd Symposium On Lapan Ipb Satellite Lisat For Food Security And Environmental Monitoring & Food Sec Envir \\
		Bioinformatics & Bioinfo \\
		Bioinformatics (Oxford England) & Bioinfo (OE) \\
		Bmc Bioinformatics & BMC Bioinfo \\
		Bmc Cancer & BMC Cancer \\
		Environmental Earth Sciences & Envir Earth Sci \\
		Environmental Modelling And Software & Envir Mod Soft \\
		F1000research & F1000 \\
		Frontiers In Psychology & Front Psych \\
		Gigascience & Giga \\
		Iop Conference Series Earth And Environmental Science & Earth Envir Sci \\
		Journal Of Pharmacokinetics And Pharmacodynamics & Pharma \\
		Journal Of Physics Conference Series & Physics \\
		Lecture Notes In Artificial Intelligence & AI \\
		Natural Hazards & Nat Haz \\
		Nature Communications & Nat Com \\
		Nucleic Acids Research & Nuc Acids Res \\
		Peerj & Peerj \\
		PloS ONE & PloS ONE \\
		Procedia Environmental Sciences & Envir Sci \\
		R Journal & R Journal \\
		Scientific Reports & Sci Rep \\
		Source Code For Biology And Medicine & Bio Med \\
		Springerplus & Springerplus \\
		Statistics In Medicine & Stat Med \\
		Studies In Health Technology And Informatics & Health Tech Info \\
		Wellcome Open Research & Well Open Res \\
		Workshop And International Seminar On Science Of Complex Natural Systems & Complex Nat Sys \\
		\bottomrule
	\end{tabular}
\end{table}

\newpage

\begin{table}[H]
	\caption{Table of abbreviations for subject tag}
	\label{Tab:Abb2}
	\renewcommand{\arraystretch}{0.4}
	\centering
	\begin{tabular}{l|r}
		\toprule
		Subject & Abbreviation \\ 
		\toprule
		Agricultural Biological Sciences &  Ag Biol Sci \\ 
		Agriculture Multidisciplinary &  Ag Multi \\
		Arts Humanities &   Art Hum \\ 
		Automation Control Systems &  Auto Cont Sys \\
		Biochemical Research Methods &  Biochem Res Meth \\
		Biochemistry Genetics Molecular Biology &  Biochem Gen Molec Biol \\ 
		Biochemistry Molecular Biology &  Biochem Molec Biol \\
		Biotechnology Applied Microbiology &  Biotech App Micro \\
		Business Management Accounting &   Bus Man Acc \\ 
		Chemical Engineering &   Chem Eng \\ 
		Chemistry &   Chemistry \\
		Communication and the Arts & Comm \& Arts \\
		Computational Biology & Comp Biology \\ 
		Computer Science &  Comp Sci \\ 
		Computer Science Artificial Intelligence &  Comp Sci AI \\
		Computer Science Information Systems &  Comp Sci Info Sys \\
		Computer Science Interdisciplinary Applications &  Comp Sci Inter App \\
		Computer Science Theory Methods &  Comp Sci Theor Meth \\
		Decision Sciences &  Dec Sci \\ 
		Earth Planetary Sciences &   Earth Plan Sci \\ 
		Education Scientific Disciplines &  Ed Sci Disc \\
		Energy &   Energy \\ 
		Engineering &  Engineering \\ 
		Engineering Electrical Electronic &  Eng Elec Elct \\
		Engineering Environmental &  Eng Env \\
		Environmental Science &  Envir Sci \\ 
		Environmental Sciences &  Env Sci \\
		Evolutionary Biology &  Evol Biol \\
		Genetics Heredity &  Genet Hered \\
		Health Care Sciences Services &  Health Care Sci Ser \\
		Health Professions &   Health Prof \\ 
		Immunology Microbiology &   Immun Micro \\ 
		Materials Science &   Mat Sci \\ 
		Mathematical Computational Biology &  Math Comp Biol \\
		Mathematics &  Mathematics \\ 
		Medical Informatics &  Med Info \\
		Medicine &  Medicine \\ 
		Medicine Research Experimental &  Med Res Exp \\
		Multidisciplinary &   Multidisciplinary \\ 
		Multidisciplinary Sciences & Multi Disc Sci \\
		Neuroscience &   Neuro \\ 
		Oncology &  Oncology \\
		Pharmacology Pharmacy & Pharm Pharmacy \\
		Pharmacology Toxicology Pharmaceutics &   Pharm Tox \\ 
		Physics Astronomy &   Phys Astro \\ 
		Psychology &   Psychology \\ 
		Public Environmental Occupational Health & Pub Envir Occ \\
		Remote Sensing & Rem Sens \\
		Social Sciences &  Soc Sci \\ 
		Statistics Probability & Stat Prob \\
		Veterinary &  Vet \\ 
		\bottomrule
	\end{tabular}
\end{table}

\bibliographystyle{biometrika}
\bibliography{kasprzak.bib}
\end{document}